**Title:** A Possible Alternative Solution to the Solar Neutrino Problem
**Author:** J.Zzimbe
**Comments:** 11 pages, PDF

The results of the Sudbury experiments have seemingly solved the solar neutrino problem. However, we must remain open to the possibility that these results may be subjected to new theoretical interpretations in the future. Should such a reinterpretation transpire, then we may find that the solar neutrino has not been solved. Should that be the case, this paper will constitute a possible alternative solution to the problem which presents new theoretical physics.

      The solution that I am proposing to the solar neutrino problem may, initially, constitute somewhat of a shock to some scientists. Consequently, although these introductory paragraphs will not state the solution, it will be alluded to in order to (hopefully) assuage the reader's (potential) shock. Any scientist worth his salt is fully cognizant of the fact that science is subject to possible modification via new discoveries. No matter how certain our theories may seem, new future discoveries may cause us to drastically alter our theoretical perspective. With this being stated, let us state the following.

      It is a self evident fact that the level of technology is affected by the level of science. The more advanced our state of science, the greater the probability of advancing our technology. What is not so evident is that the state of science in one area can potentially be affected by the degree of advancement in another area of science. An example to illustrate this is as follows. Let's work within the parameters of a hypothetical scenario and assume we are oblivious to the following scientific facts. Namely, $E = mc^2$ and nuclear reactions (specifically, nuclear fusion). If this situation prevailed, what would be our contemporary theoretical perspective about the source of stellar energy? Presumably, we would still advocate a theory which we once believed in. Specifically, the gravitational theory proposed by Hermann Helmholtz. This theory entails that stars are gradually shrinking in diameter. An incremental decrease in volume was believed to account for the energy emitted by a star. In the case of the sun, if its diameter shrank by 200 feet per year, this would explain its energy output. However with the development of new scientific principles, ($E = mc^2$ and fusion) we were able to realize that it was fusion which provided an energy source for stars. Up to this stage in the development of theoretical physics, there is absolutely no reason to adopt the position that fusion is not the primary source of stellar energy. However, this paper may cause that tenet to be seriously questioned.

      The reader may have some rudimentary concept of what this paper will propose as the solution to the solar neutrino problem. If the reader has made the correct inference, there may already be a substantial amount of shock on the part of the reader. To be specific, this paper may be analogous to Pauli's proposal of the neutrino. A radical concept is being proposed *merely* to solve a significant scientific problem. This is not the case. In order to illustrate this, here is a *brief* explanation of the genesis of this theory.

      When I started studying physics, I was "discontent" with the lack of development of a certain theoretical framework within physics. I felt a dire need for the continued development of an aspect of physics. Upon completing the analysis, I came to the realization that the principles I had developed must be extrapolated to stellar dynamics. It was subsequent to this that I realized that my framework could offer the possible solution to the solar neutrino problem. Therefore, these

new concepts were not developed with the *intent* of solving the solar neutrino problem. I was merely seeking a deeper comprehension of a certain aspect of physics. The deeper comprehension I was seeking is explained via the following.

When the wave-particle duality of light was established in the early 1900's, it was, of course, a step forward for physics. However, a deeper analysis is required with respect to the nature of light. To state that light is comprised of waves of light or photons of light is grossly inadequate. If one is trying to define light, one cannot utilize the same word (light) within the definition of the word that one is trying to define. What is the *precise* nature of the waves, and what is the *precise* nature of the photons? The reader may feel that an analysis of this nature will inevitably lead to a physics catastrophe of one kind or another as a result of the historical precedence established in the late 1800's. Specifically, the postulation of the ether and its subsequent failure. Therefore, there should be clarification of the position being advocated. It is *not* being stated that light cannot propagate in a vacuum. Nor will it be stated that there is a requirement of some type of medium for light to propagate in since the sun's light reaches us via the vacuum of outer space. There is no doubt that light does not require a medium in which to propagate. Light propagates very well through a vacuum without the need for some type of medium. What is being advocated is a more intense analysis that provides a much deeper comprehension of the nature of waves and photons. In this paper, light will be defined as follows.

> *Light is comprised of waves. These waves are comprised of photons. Each individual photon propagates with a sinusoidal motion.*

A quantitative analysis to describe the motion of a singular photon is as follows. Firstly, it must possess a relativistic velocity. This is established via the following.

$$x = ct \tag{1}$$

Its sinusoidal motion would be characterized via the following.

$$y = a \cos(w_o t + f) \tag{2}$$

By expressing equation one via $t$, (namely $x/c$), the equation to describe a photon is expressed by modifying equation two.

$$y = a \cos\left(\frac{w_0 x}{c} + f\right) \tag{3}$$

When dealing with a photon, the "factor" (its precise nature will be outlined) which induces its velocity (represented by $c$ in equation one) and sinusoidal motion are intricately connected with each other. Namely, the "factor" which induces a photon's relativistic speed also induces its sinusoidal motion.

In order to increase our comprehension of light, the following should be implemented. Firstly, the precise nature of photons must be analyzed. From there, it would be most judicious to analyze how the photons "conglomerate" in order to propagate as cohesive waves. For reasons that will subsequently be stated, only the former (analyzing the precise nature of photons) will be analyzed in this paper. However, analyzing the precise nature of photons will suffice to provide the reader with the new theoretical physics required to solve the solar neutrino problem.

The primary experimental facts to be explained via a theoretical analysis is why the photon is relativistic and why it has a frequency associated with it. The simplistic answer to this is to state that the energy within the photon exists in a certain state so as to imbue the photon with a relativistic speed and a frequency. An explanation behind this nebulous statement is as follows.

In order to provide some degree of lucidity to the explanation, certain spatial orientations will be attributed to the photon. A very loose representation of the photon is as follows.

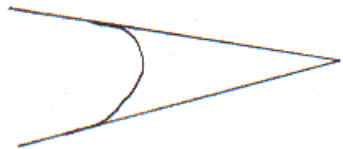

Fig.1

A representation of this nature has been developed for more than one reason. Firstly, this representation is successfully indicative of a spin one particle. Since it has a superficial resemblance to an arrow, it would have to spin 360 degrees before it looked the same way again. However, there are more cogent reasons for a representation of this nature.

An (approximate) representation of this nature would serve to explain what was previously stated. Specifically, that the energy within the spatial confines of a photon exists in a certain state so as to imbue the photon with a relativistic speed and frequency. To begin, we shall analyze why this type of representation would induce a relativistic speed.[*]

As the reader can see, towards the back end of the photon there is, what could potentially be labelled, a type of "hole". In other words, there is no "back end" to the photon. The reason this would result in a relativistic particle is as follows. The energy within the photon could not be "evenly distributed" throughout the photon as a direct result of the unusual three dimensional configuration of the photon. Since the photon is, in essence, devoid of a "back end", the energy would not be effectively diverted there. However, since the representation provided is indicative of a "front end" that comes to a point, the photon's energy would be concentrated and directed towards the front of the photon. The fundamental ramifications of this are as follows. All of the photon's energy is concentrated in one direction within the photon.  Since that direction is towards the "front end", this would inevitably induce a relativistic speed to the photon, assuming the energy is sufficient in magnitude.  That energy would obviously be sufficient in magnitude as all photons are relativistic. (The reader may object to the argument presented when it is compared with the pictorial representation provided.  The pictorial representation would not necessarily be indicative of energy being directed towards the front, even if there was no "back end" to speak of.  There are other areas within a photon where the energy could potentially be diverted.  There are two things to be stated in relation to an argument of this nature.  Firstly, the frequency of the photon, and its relation to this configuration, has yet to be explained.  Secondly, the pictorial representation is only meant to be an approximation.  This is inevitable as nobody has ever "seen" a photon.  The representation is only meant to be indicative of the following.  The photon is "configured" in such a way so that the photon's internal energy would be concentrated and directed towards the front.  A particle that was devoid of a "back end" but came to a point at the front would suffice to achieve this goal.  The pictorial representation provided conforms to these broad parameters.)

---

[*] In the following, it is acknowledged in advance that some of the terms in quotes do not fall within the parameters of appropriate physics terminology. Regrettably, I don't believe appropriate terminology exists for the aspects I am striving to explain.

      The second quality of the photon that needs to be explained is its frequency and the reason that photons possess different frequencies.  The frequency of the photon can simply be attributed to the photon propagating with a sinusoidal motion.  But, the question is raised as to what *induces* this sinusoidal motion?  An explanation is as follows.  The photon's energy could not be continually directed towards the front unless there was some reason the energy were directed in a backwards *direction* (as opposed to the back end) and redirected towards the front.  The latter process just explained would account for the photon's frequency.  The specifics are as follows.

      Immediately subsequent to energy being directed against the front of the photon, that energy will scatter in a backwards direction that is either in a downwards or upwards motion.  What "factor" will decide whether the energy proceeds upwards or downwards is unknown.  It may be nothing more than a fifty-fifty chance.  Let's assume the energy proceeds upwards.  When it initially proceeds in the backwards/upwards direction, once it scatters into the top end (towards the back) of the photon this would be the equivalent of the photon's "crest".  (Even though the word crest is properly applied to wave phenomena as opposed to particles, we will nevertheless utilize the word crest in this context).  That same energy would then scatter from the crest *straight downwards* into the bottom end in order to establish the photon's "trough".  From there, the energy would then be directed in a forwards motion again.  It was stated that there may be nothing more than a fifty-fifty chance as to whether the energy is directed upwards or downwards once it begins to propagate in a backwards direction. Regardless of what may potentially decide this direction, the polarization of the photon will be determined by the initial direction in which the energy flows.  According to *The Feynman Lectures on Physics* volume I, chapter 33, page 2, light which proceeds in an anticlockwise direction will have right-hand circular polarization.  If the energy is initially directed upwards (and towards the back of the photon), the photon will have right-hand circular polarization. Should this explanation be lacking lucidity, an analogy will elucidate.  Let's assume we have a pencil sitting on a table with the lead tip pointing in the right direction.  If we push the lead tip upwards, the pencil will go upwards/anticlockwise.  (This is not the best analogy as the energy within a photon is directed in a backwards *direction*, whereas the pencil is being pushed upwards at the front.  Notwithstanding this fact, this analogy will have to suffice for our purposes.) This is the equivalent of right handed circular polarization with a photon.  Conversely if the energy is initially directed downwards (and towards the back), the photon will possess left-hand circular polarization.  (If the *initial* direction of the energy is downwards, the propagation of the photon will be initiated at the trough).  Therefore, the photon's sinusoidal motion/frequency and polarization

has been explained. (The photon's circular polarization has not yet fully been explained. There are certain critical details which must be provided in order to complete the explanation. However, for reasons that will eventually be made clear, the explanation of circular polarization will not be completed in this paper despite the fact that I have completed a theoretical analysis.)

The next question may naturally be, what is the difference between a red photon and a violet photon. The simplistic answer is that a violet photon possesses more energy than a red photon. However, more must be stated. It is my opinion that the primary difference amongst the energy levels of photons is not in the amount of energy contained therein, but rather the same energy is being scattered within a smaller physical area. An explanation (and analogy) is as follows. Let's take the Heisenberg uncertainty principle. Let's assume we have a particle in a box. If we "magically" reduce the size of that box, the particle will reflect from the walls with greater "vigour" as its momentum has been increased. This is superficially similar to the nature of photons. A more energetic photon is a photon which is in possession of the same amount of internal energy as a less energetic photon, but has smaller dimensions. In the Heisenberg uncertainty principle, it is equivalent to a particle being in a small box or the same particle being in a larger box. A red photon which is larger than a violet photon is *perceived* as possessing less energy as the intrinsic energy is distributed over a larger area. More precise details are as follows. If a red photon is in possession of slightly larger dimensions than an orange photon, the internal energy would have a slightly greater distance to traverse when propagating in a backwards direction before it initially scattered from the (let's say) top of the photon. At that stage, there would be a greater distance to traverse before it scattered into the bottom of the photon (in relation to an orange photon). *As a direct result of these greater differences* one photon would have a lower frequency than another photon. The reader may feel that there is a substantial flaw in this theory. If this concept of greater differences in the area to be traversed were true, since a red photon has a measurable difference in its frequency, then there should also be a measurable difference in its velocity. Yet both the red and violet photons are relativistic. This precept does not invalidate the theory for the following reason. The majority of the photon's intrinsic energy is directed in a forwards direction with only a limited amount "designated" (this is a loose term) for the photon's frequency. Therefore, the energy is *perpetually* impacting in the forwards direction. In order to achieve this constant impaction, there is (and this is speaking in extraordinarily loose terms), a "line up" of energy towards the front of the photon. In other words, there is a conglomeration of energy towards the front which serves to achieve this constant impaction. This is in stark contrast to the energy which is (again speaking loosely) "sporadically" propagating to induce the photon's frequency.

Now that there is a somewhat complete analysis of the photon (I am cognizant of the fact that light being comprised of waves has not been fully addressed), new physics can begin to emerge. According to Feynman, volume I, chapter 42, page 9, it was, "... Planck's viewpoint that only oscillators of matter were quantized,... ". It was Einstein who extrapolated this to light and ended up playing the most important role in establishing the photon theory of light (at least in my opinion). The equation $E = fh$ is applied to energy in general and shows that energy is quantized. What has been established thus far in this paper, produces an equation that is somewhat similar to Planck's quantum theory but is *only* applicable to light. This paper provides a more in depth analysis of light by examining the "internal mechanism" of light. It shows that light is a *source* of energy, not merely a form of energy. If light is subjected to "certain forces" (which will be momentarily described), light can be broken down to function in the capacity of an energy source. An equation to characterize the overall energy of light is as follows.

$$E = \sum_{\text{Foch space}} fh(\text{CONSTANT}) \qquad (4)$$

The constant will be the precise value of the *internal* energy of a photon (which will be the same regardless of the photon's frequency). Since this equation is so similar to Planck's quantum theory, the reader may wonder why we should even bother with it. Although the equations are similar, this paper is presenting a slightly more developed theory of light. The internal energies of all photons are the same. However, if that same energy is distributed over a smaller physical area, there will be certain discernible ramifications. Planck's quantum theory does *not* state that the internal energies of all photons are the same.

Given what is being postulated, it may be appropriate to point out the following. If we wish to view light in the capacity of a *source* of energy (as opposed to a mere form of energy) then the term $(fh)$ would be fundamentally superfluous. This is for the simple reason that if we wished to assess the *internal* energy of all the photons within a designated Foch space, the energy of every photon would be the same. Therefore, a more appropriate equation may be as follows.

$$E = I_{(c)} \text{ (CONSTANT)} \qquad (5)$$

The term $I_{(c)}$ simply refers to the intensity of light (i.e. the number of photons). However, the difference between equations four and five may be nothing more than a question of semantics. On the one hand a violet photon seems to be more energetic than a red photon, yet on the other hand they both possess the same amount of internal energy.

The genesis of special relativity was Einstein asking himself what he would see if he were to ride a beam of light. In other words, a physics principle (linear motion) was "taken to an extreme". If a particular principle outlined in this paper were "taken to an extreme" would new physics emerge? What is specifically being referred to is the consideration of a photon having its frequency increased to a phenomenally high level (i.e. beyond the energy levels of gamma rays). New physics does emerge which is of the following nature.

As the frequency increases, the spatial dimensions of the photon becomes smaller. If the frequency were to increase to a very high level, the dimensions would become so small, that the energy could no longer be contained within the photon. The energy would, for all intensive purposes, explode from within the confines of the photon. The photon would no longer exist but would be converted to energy. The reader may feel that this is an interesting concept in principle. However, in practical terms an "exploding photon" could not exist under any circumstances. As an example, a red photon cannot be "injected" with a sufficient amount of energy to cause it to "explode". This is not necessarily the case. General relativity dictates that as a photon proceeds towards a gravitational field, its energy will increase. (As a result of this paper, that entails that its spatial orientation has become smaller). Therefore, if a gravitational field (*or the equivalent thereof*) were sufficient in magnitude (perhaps a black hole) then a photon propagating directly towards it may reach sufficient energies to cause the photon to "explode".

This is the new theoretical physics which will enable us to solve the solar neutrino problem (once it is incorporated into certain aspects of stellar dynamics). The reader may feel that this analysis is grossly inadequate for the following reason. Earlier in the paper, it was stated that a complete analysis of light requires an explanation as to how the photons "conglomerate" in order to propagate as cohesive waves.  There is no question that the analysis of light is incomplete until this theoretical aspect is fully addressed.  However, it is unacceptable to address this aspect in this paper.  The precise rationale behind this is as follows.

      Since this archive is not refereed, there are certain responsibilities which are expected of submitters.  There is an onus on the submitter to ensure that the paper will not, in one respect or another, waste the reader's time.  The factor that is germane to this paper pertains to the presentation of radical concepts.  The precise coupling mechanism that I have postulated is of a *highly* radical nature.  It is fundamental human nature to be reluctant to accept anything of a radical nature.  The same is true of scientists. Most scientists would be vehemently opposed to the coupling mechanism I have postulated.  Given this fact, I will not provide my complete analysis despite the fact that I have written an entire paper on the nature of light.  Furthermore, although this analysis of light is incomplete without the theoretical explanation of how individual photons "conglomerate" to propagate as cohesive waves, the analysis of photons is sufficient to proceed with a possible solution to the solar neutrino problem.  We will now return to the solution to the solar neutrino problem.

      Since a new theory has been presented adopting the position that light is not only a form of energy but also a source of energy, this new theoretical aspect must be incorporated in to stellar dynamics.  Is it possible that light itself could be responsible for generating at least a portion of a star's energy output?  The very existence of the solar neutrino problem would provide some evidentiary foundation to indicate that this position is tenable.  When calculating the number of neutrinos that the sun should emit, we have, of course, worked on the assumption that fusion was the primary/only source of stellar energy.  We had no choice in adopting this position as our scientific framework did not indicate that there was another viable source of solar energy.  However, if a viable alternative source of stellar energy is outlined, then this may solve the solar neutrino problem.  Namely, there are fewer neutrinos being emitted by the sun than theory predicts as a result of fusion functioning in the capacity of a secondary source of stellar energy.  There is another source of energy which is the primary source for stellar energy.  This other source would be light, or more specifically, photons (as opposed to entire waves of light).

      The reader may be wondering how the principles previously outlined in this paper (as they pertain to photons) could possibly be extrapolated to stellar dynamics in such a manner that photons would constitute a source of energy for stars.  To briefly recapitulate, it was stated that photons could no longer contain their intrinsic energy if the frequency of the photon became phenomenally high.  Specifically, the spatial dimensions of the photon would become sufficiently small that the energy could no longer be contained within such small dimensions and the photon would, for all intensive purposes, explode, thereby releasing its internal energy.  In accordance with general relativity, a photon might achieve this state if it were subjected to an enormously strong gravitational field, or the equivalent thereof (thereby dramatically increasing its frequency).  The reader may feel that this last sentence would automatically render this entire concept (photons constituting the primary source of stellar energy) useless as there is no scenario under which photons, within the sun (or any star), could be subjected to the requisite forces to achieve this effect.  The rest of this paper will outline how it is possible for this scenario to take effect and what pre-conditions must be in effect for a photon within the sun/star to "explode" and release its energy.

      For the sake of the physicists who may be reading this (and have not taken an astronomy course for a substantial period of time), one small aspect of astrophysics will be briefly reviewed.  Within a stellar interior, an emitted photon may only travel the tiniest fraction of a centimetre

before it is re-absorbed. A low frequency photon would not have a sufficient "opportunity" to be subjected to the requisite forces (whose precise nature will subsequently be outlined) to enable it to gain a sufficiently high frequency to induce the previously described effect. Therefore, the photon, *at the time of its emission* must fall within the following parameters. The photon must be ultraviolet and of *very high* frequency. This latter sentence introduces the question, precisely what *kind* of photon would this be? This is not an easy question to answer with precision given the fact that the energy levels between X-ray photons and gamma ray photons can, at times, somewhat "overlap". As an example, it is possible that a high energy X-ray photon could potentially be labelled a low energy gamma ray photon, and vice versa. Therefore, for the time being, it will suffice to state that the photon may be a high energy X-ray photon or a low energy gamma ray photon.

    At what general area of the star is there the highest probability of achieving photons that conform to these parameters? At the core of the star. The rationale behind this is as follows. When dealing with free electrons, the frequency of an emitted photon is directly proportional to the acceleration of an electron. Since it is hottest at a star's core, the magnitude of an electron's acceleration would be greatest at the core. Consequently, the following has been established. It is hottest at a star's core. This is where the magnitude of acceleration for free electrons would be greatest. Therefore, it is at a star's core that there would be the highest probability of an inordinate number of high frequency photons being present (even if these photons are present only for a fraction of a second before they are absorbed).

    The reader is still vehemently opposed to considering the possibility of photons constituting any kind of energy source for the sun (let alone the primary source) since at the sun's core there doesn't exist the requisite gravitational force to increase a high energy photon's frequency any further. Although this, of course, is true, the reader must give due consideration to the following. It would require a gravitational force, *or the equivalent thereof* to increase a photon's frequency. General relativity's principal of equivalence states that an object being subjected to an acceleration is the equivalent of an object being subjected to a gravitational field. The reader is no doubt still opposed as a preliminary analysis indicates that photons could not be accelerated. A *preliminary* analysis is indicative of this. A deeper analysis dictates the following.

    The sun is in a state of constant motion in various respects (the following analysis will apply to other stars as well). Firstly, the sun, like the earth, is rotating on an imaginary axis. Secondly, the sun is orbiting the galactic centre. Thirdly, the entire galaxy (and consequently all stars in that galaxy) is moving at a very rapid rate. Therefore, there are at least three different states of motion that the sun is constantly subjected to. The reader is still opposed to the concept of photons being accelerated within stellar interiors as there is no acceleration involved in any of these states of motion. The rotation about the axis, the orbiting of the galactic centre, and the motion of the entire galaxy are all constant. (The only exception would be the orbiting of the galactic centre since there would be times when, like the planets of the solar system, this orbital rate is slightly faster or slower. Nevertheless, in all probability, this particular motion would not possess the requisite rate of acceleration). This, of course, is true. However, there is a critical element that has been overlooked in a criticism of this nature when considering the possibility of "exploding" photons within stellar cores.

    Since the principal of equivalence was originally predicated upon the concept of an elevator in outer space, we shall adhere to this example in order to illustrate the desired precept. We only feel an acceleration when an elevator *initially* starts. Subsequent to that initial start, the elevator is in constant motion. Let's assume that the elevator is moving upwards in a state of constant motion

and the following hypothetical scenario takes effect. A ball "magically appears" in the middle of the elevator. The ball would hit the floor as the elevator proceeds upwards. This would be the equivalent of an acceleration/gravitational field. Therefore, if an entity "appears" within a second entity that is in a state of constant motion, even though the second entity is not accelerating but, instead in a state of constant motion, the first entity would be subjected to the equivalence of an acceleration by virtue of its "appearance". Furthermore, if a certain action is equivalent to an acceleration, this would be equivalent to a gravitational field.

Photons do appear within stars. When an electron is accelerated within a star, a photon will be emitted. If certain conditions prevail, the photon will be subjected to an acceleration (by virtue of the various states of motion a star is subjected to) and its frequency will be increased. The pre-conditions which must prevail in order to achieve a state whereby a photon releases its energy are as follows.

Firstly, as has already been stated, the photons must be ultraviolet and of very high frequency. Therefore, our analysis will be implemented via high energy X-ray photons and low energy gamma ray photons. (It goes without saying that high energy gamma ray photons would also conform to the parameters outlined in the following analysis). Secondly, the various states of motion a star is subjected to must be properly analyzed. (In the following equations, the velocities are to be computed via a vector addition).

$$v_{(\text{galaxy})} + (\mathbf{w} \cdot r)_{(\text{of an individual star})} = v_{(\text{final of an individual star})} \qquad (6)$$

Equation number four is applicable to stars regardless of their precise orbit around the galactic centre. To be specific, equation four is germane regardless of whether the star's orbit is perfectly circular (a steady orbit), or whether the star's galactic orbit varies via a smaller or greater radius. The second term, $(\mathbf{w} \cdot r)$, ensures that both types of motion are incorporated into the equation.

In the above equation, the reader may feel that certain critical factors have either been omitted, or not given appropriate consideration. Firstly, the reader may feel that the star's rotation about its own axis has not been taken into consideration. Although I am cognizant of a star's rotation about its own axis (as was stated in the previous page), this has not been factored into the equation for the following reason. A star's rotation about its own axis is minimal *when compared* to the velocity of the galaxy, or a star's orbital velocity about the galactic centre. Consequently, it would have virtually no influence on the final effect of a photon releasing its final energy and is not incorporated into the equation. (The only justification for including the star's rotation about its own axis, would be if one is striving to ascertain the final sum in equation four to several decimal places).

The second factor which the reader may feel has not been given appropriate consideration pertains to the velocity of the galaxy. What is specifically being referred to is the cosmological constant. The reader may feel that as a result of the cosmological constant, the velocity of the galaxy should be expressed in the equation. However, this has not been implemented for the following reason. I don't feel that it's appropriate to introduce one aspect of science (specifically, the cosmological constant) until we are far more certain about the accuracy of this scientific principle. Furthermore, before incorporating the cosmological constant, we should gain a greater comprehension of the rate of acceleration and at what times this acceleration takes place (assuming of course that the universe is actually accelerating). To express it succinctly, our knowledge and comprehension of the cosmological constant is far too simplistic to effectively incorporate into the equation. Consequently, a galaxy's velocity has not been expressed with consideration to the cosmological constant.

The foundation for a photon to release its energy within the core of a star (and subsequently function in the capacity of the primary source of stellar energy) has now been established. Firstly, a star's final velocity must be ascertained via equation four. Secondly, an abundance of high energy photons (either high energy X-ray photons or low energy gamma ray photons) must be continually emitted within the star (as already stated, this would obviously be at the core of the star). If the requisite photon is emitted in a direction *opposite* to the *final state* of motion of a designated star, its frequency will be increased. Or, to express it in terms of the precepts previously stated in this paper, the photon's dimensions will be reduced. Consequently, the internal energy can no longer be contained within this reduced area, thereby causing the photon to "explode" and liberate its energy.

Now that the preconditions for a photon to release its energy has been established, it would be appropriate to incorporate another factor into this analysis. A photon would not necessarily have to be of a very high frequency *at the time of its emission* in order to (eventually) have its dimensions reduced and its energy released. This can be explained via the following.

According to the tenets outlined in this paper, a violet photon (or for that matter, a low energy X-ray photon) would not have its energy released at the time of its emission even if it were travelling in a direction opposite to the final velocity (as computed via a vector addition) of the star. This is for the simple reason that it will be reabsorbed within a fraction of a second of its emission and will not have the requisite time to have its dimensions sufficiently reduced. However, in accordance with high frequency photons having their dimensions reduced, a violet photon (as an example) *would* have its dimensions reduced if it were emitted in a direction opposite to the final velocity of the star. Since its dimensions are reduced, its energy would increase (in accordance with the previously stated precepts) and it may become a low energy X-ray photon. If this process were to transpire again, it may then "acquire" the requisite energy to "explode" and liberate energy. Consequently, photons which lacked the requisite energy levels to release their energy at the time of their emission could potentially have their frequency increased in order to function in the capacity of an energy source. This, of course, is contingent upon the photons being subjected to the necessary forces. To take this analysis slightly further, in *principal*, even an infrared photon could (eventually) achieve the requisite energy levels if it was successively emitted in this manner. However, this is true only in principal. The actual probability of such an event is quite low. The probability of this process is far greater with violet photons or low energy X-ray photons. Consequently, not only would high energy X-ray photons and gamma ray photons (regardless of whether the gammas are high energy or low energy) function in the capacity of a source of stellar energy, but violet photons and low energy X-ray photons would function in the same capacity *if* they were *successively* subjected to the forces outlined in this paper.

Although the mechanism of a photon constituting the primary source of stellar energy has been outlined, other principals should be briefly addressed. Photons functioning in the capacity of an energy source could only be expected within the core of a star (barring mankind's ability to dramatically increase our level of scientific understanding and our technological capabilities). The rationale is as follows. As the reader should be able to see from the preceding analysis, various factors of a highly precise nature must come into play for a photon to release its energy. At the exact time of its emission, the photon must be of very high frequency and its direction of propagation must be *directly opposite* to the star's final state of motion (which, as already stated, is dictated via a vector addition of the star's orbit around the galactic centre and the galaxy's overall state of motion). The probability of this transpiring with any degree of regularity is extremely low unless there are an inordinate number of high energy photons within the designated area. When photons are emitted, they could potentially assume any direction of propagation. Only a tiny fraction of them would propagate in a direction exactly opposite to the star's final state of motion. Therefore, the only place where there would be a *truly inordinate* number of high energy photons being released is within the core of a star where it is hottest and free electrons are constantly being accelerated to a very high speed. Although it would be possible for this to transpire anywhere within the star, the probability of it transpiring in other areas (at least with any degree of regularity) is extremely low. The second principal that should be addressed is that fusion is not to be eliminated as a source of stellar energy. It is merely to be relegated as a secondary source of stellar energy, not the primary source.

     As was stated at the beginning of the paper, should the results of the Sudbury experiments be reinterpreted, then this paper will provide a possible alternative to the solar neutrino problem. Furthermore, the solution is achieved via new theoretical physics.